\documentclass{article}

\usepackage{PRIMEarxiv}

\usepackage[utf8]{inputenc} 
\usepackage[T1]{fontenc}    
\usepackage{hyperref}       
\usepackage{url}            
\usepackage{booktabs}       
\usepackage{amsfonts}       
\usepackage{nicefrac}       
\usepackage{microtype}      
\usepackage{lipsum}
\usepackage{fancyhdr}       
\usepackage{graphicx}       
\usepackage{tcolorbox}
\tcbuselibrary{breakable}
\graphicspath{{media/}}     

\title{Repositioning Tiered HotSpot Execution Performance Relative to the Interpreter
\thanks{\textit{\underline{}}:
}}

\author{
  Jonathan Lambert \\
  Department of Computer Science \\
  Maynooth University \\
  Maynooth, Co. Kildare\\
  Ireland\\
  \texttt{jonathan.lambert@ncirl.ie} \\
   \And
  Kevin Casey \\
  Department of Computer Science \\
  Maynooth University \\
  Maynooth, Co. Kildare\\
  Ireland\\
  \texttt{kevin.casey@mu.ie} \\
  \And
  Rosemary Monahan \\
  Department of Computer Science \\
  Maynooth University \\
  Maynooth, Co. Kildare\\
  Ireland\\
  \texttt{rosemary.monahan@mu.ie} \\ 
}

\begin{document}
\maketitle

\begin{abstract}
Although the advantages of just-in-time compilation over traditional interpretive execution are widely recognised, there needs to be more current research investigating and repositioning the performance differences between these two execution models relative to contemporary workloads. Specifically, there is a need to examine the performance differences between Java Runtime Environment (JRE) Java Virtual Machine (JVM) tiered execution and JRE JVM interpretive execution relative to modern multicore architectures and modern concurrent and parallel benchmark workloads. This article aims to fill this research gap by presenting the results of a study that compares the performance of these two execution models under load from the Renaissance Benchmark Suite. This research is relevant to anyone interested in understanding the performance differences between just-in-time compiled code and interpretive execution. It provides a contemporary assessment of the interpretive JVM core, the entry and starting point for bytecode execution, relative to just-in-time tiered execution. The study considers factors such as the JRE version, the GNU GCC version used in the JRE build toolchain, and the garbage collector algorithm specified at runtime, and their impact on the performance difference envelope between interpretive and tiered execution. Our findings indicate that tiered execution is considerably more efficient than interpretive execution, and the performance gap has increased, ranging from 4 to 37 times more efficient. On average, tiered execution is approximately 15 times more efficient than interpretive execution. Additionally, the performance differences between interpretive and tiered execution are influenced by workload category, with narrower performance differences observed for web-based workloads and more significant differences for functional and Scala-type workloads.
\end{abstract}

\keywords{Java Runtime Environment \and Java Virtual Machine \and JRE Build Toolchain \and GNU GCC \and Tiered Execution \and Tiered Just-in-time Execution \and Interpretive Execution \and HotSpot \and C1 Compiler \and C2 Compiler \and Renaissance Benchmark Suite \and Raspberry Pi 4 Model B.}

\section{Introduction}
Over the last 30 years, the Java programming language and its associated runtime environment have gained widespread acceptance and is well-positioned to maintain its popularity \cite{Belani}. Initially designed as a platform-independent language with enhanced security features through its managed memory model, it has become the target for many other languages seeking to leverage those same characteristics. However, the language was initially criticised for its poor execution time performance, particularly compared to compiled code \cite{Gu, Aycock, Prechelt}. Poor execution time performance was especially true for interpreted Java bytecode \cite{Cramer1, Hsieh}. To address this issue, early in its development, the language integrated a just-in-time (JIT) compiler (HotSpot) and offered two variants through its client and server architectures.

In recent years, Java has transitioned to a tiered model of execution \cite{OracleTierdJVM} that integrates the benefits of interpretive execution, the fast compilation optimisations associated with the client compiler, and the heavier optimisations related to the server compiler. Early evaluations of the JIT compilation of Java bytecode and the subsequent execution of native code showed significant performance improvements over solely interpreted execution \cite{Cramer1}. Over the past three decades, the number and complexity of JIT optimisations have considerably increased, moving from local-level optimisations within methods, for example, constant folding, common sub-expression elimination, method inlining, null-check elimination, to optimisations identified from analysis that is considered more expensive and resource intensive, such as techniques that rely upon escape analysis that have a global-level effect, for example, stack allocation and dead object elimination, synchronisation elimination if an object does not escape its current thread, or array bounds check elimination if an array is accessed only within its bounds.

Over the last three decades, Java code has been deployed and executed on various computer architectures, with the JRE and associated JVMs targeting ARM 32 and 64 bit architectures, Intel x86 and x64 architectures, and SPARC 64 bit architectures, for example. Modern multi-core CPU architectures have provided Java with access to resources that have widened its performance envelope, particularly with respect to multi-core parallelism. This access has allowed Java's memory management model to considerably increase its efficiency by integrating parallel garbage collectors and enabling the C1 and C2 compilers to work in parallel with the main execution thread.

The relationship between interpretive performance and tiered execution performance can be viewed from a dependency perspective, in that interpretive execution is the first tier of execution within the tiered execution model. With that said JVM interpretive execution can be the sole mode of execution and is of importance from a number of additional perspectives above and beyond its role within tiered execution. For example, the availability of JVM interpreters allows languages to target architectures without the need to construct dedicated compilers, leveraging decades of research and significantly reducing the costs associated with modern language development. Interpreters can substantially facilitate the development of modern languages, new features, and the testing of alternative paradigms. In addition, interpreters are more adaptable than compilers \cite{Proebsting}, offering simplicity and portability \cite{Gagnon}, compact code footprint, significantly reduced hardware complexity, as well as substantial reduction in their development costs \cite{Maierhofer}. Finally, tasks such as debugging and profiling can be accomplished easier through an interpreter compared to a compiler. 

This paper examines the performance gap between interpreted Java execution and tiered execution. Specifically, it investigates the performance gap between these two modes of execution relative to a contemporary collection of Java workloads that stress the JVM with modern concurrent and parallel tasks. Furthermore, we consider those performance relative to an array of OpenJDK Java Runtime Environments (JREs), ranging from version 8 through to version 14. In addition, to reposition the performance gap relative to a contemporary collection of workloads and modern CPU architectures, this paper also explores whether build toolchain compiler choice impacts overall JRE performance. 

The main contributions of this work are:

\begin{itemize}
    \item A systematic analysis of the performance gap between interpretive and tiered execution;
    \item Analysis of the performance gap between interpretive and tiered execution relative to a contemporary modern collection of concurrent and parallel workloads;
    \item Observations of the widening performance gap between interpretive and tiered execution;
    \item Recognition that workload characteristics are a significant factor in determining the degree to which the performance differs between interpretive and tiered execution;
    \item Out-of-box default garbage collector parameters have no impact on performance differences between interpretive and tiered execution, supporting the need to specifically tune the garbage collector parameters for the task at hand;
    \item Evidence that build toolchain compiler choice has no impact on performance.
\end{itemize}

The paper is structured as follows, in Section \ref{sec2}, we present an overview of the background associated with Java bytecode performance, and we follow that in Section \ref{sec3} with a detailed exposition of the methodology followed concerning the capturing of all measurements. In Section \ref{sec4}, we present the results of this study, and in Section \ref{sec5}, we give our discussion of those results and position our findings relative to the literature. We finally present our conclusions, limitations, and future work in Section \ref{sec6}.

\section{Background and Related Work}\label{sec2}
The relationship between Java interpretative execution and just-in-time execution performance relative to each other is still of contemporary relevance, not just because of their integrated role within the modern JVM and specifically tiered execution \cite{Polito} but also concerning the extent to which it is relied upon for the execution of code on a vast array of platforms, including a myriad of embedded devices. 

The advantages of targeting the JVM include the ability to run on lightweight, small-footprint JVM implementations targeted at embedded devices \cite{Gough}. A JVM interpreter has the characteristics to run on embedded devices, notably due to their small memory footprint and their reduced need for dynamic memory. This is in contrast to JIT compilers that require additional resources to optimise the code, typically in parallel, while the JVM continues interpretive execution. Even when space is not a constraint, using interpreters in the translation of cold methods, especially for short-running applications, helps reduce compilation overhead \cite{Azevedo}. In addition, the contemporary JVM is a popular target for many languages, primarily because of its ability to provide an environment to efficiently and competitively execute bytecode, for example, Scala, Groovy, Clojure, Jython, JRuby, Kotlin, Ceylon, X10, and Fantom \cite{Urma}. The availability of JVM interpreters allows languages to target architectures without the need to construct dedicated compilers, leveraging decades of research and significantly reducing the costs associated with modern language development. 

Both interpretive execution and just-in-time execution have seen a large body of research focus, possibly more in recent years, toward just-in-time execution rather than interpretive execution. In regards to interpretive execution, early work had considered the dynamic semantics of interpretation and identified that the interpreter dispatch loop was primarily at fault and contributed significantly to the JVM's underperformance when executing in interpretive mode. Four dispatching mechanisms are prevalent and relied upon for implementing Java interpreters. The techniques for instruction dispatch that have been prominent in implementations have been: switch-based, direct threading, direct call threading, and inline direct threading \cite{Casey, Gagnon}. The switch-based approach is the simplest but the least efficient, requiring a minimum of three control transfer instructions for each instruction.

In contrast, direct threaded dispatch updates the internal opcode structure with the address of the implementation, eliminating the instruction switch and encapsulating the dispatch loop. Direct threaded code's instructions control the program counter while the switch expression updates it. Inline threaded implementations amalgamate instruction sequences within bytecode basic blocks, removing dispatch overhead associated with control transfer instructions for all instructions except the last one. Casey et al. \cite{Casey} and Gagnon et al. \cite{Gagnon} provide comprehensive overviews of all three techniques. Early research identified instruction dispatching as the most expensive bottleneck for bytecode execution. 

Apart from the interpreter optimisations previously mentioned, the modern JVM interpreter, known as the "template interpreter," has taken a different approach to instruction dispatch at a lower level. The template interpreter achieves this by creating a dispatch table that maps Java bytecode to optimised assembly instruction sequences. During the build process, the template interpreter's construction enables the JVM to recognise the architecture it is running on, producing optimised code for each bytecode, called codelets \cite{Ritter}. These codelets are designed for individual bytecode instructions and frequently executed bytecode sequences.

According to Simonis' research \cite{Simonis}, as cited by Ritter in \cite{Ritter}, who undertook a performance comparison between the template interpreter and the historical C++ interpreter, the template interpreter outperforms the C++ interpreter by a factor of two to three. The interpreter is an integral component within mixed-mode execution. Simonis \cite{Simonis} also considered the difference in performance between mixed-mode execution with a template interpreter versus mixed-mode execution with a C++ interpreter. Their results showed a differential in favour of mixed-mode execution with a template interpreter. Mixed-mode execution performance with the template interpreter versus the C++ interpreter ranged from 1.1 times to twice as fast. 

Focusing on the just-in-time compilers composing the JRE and its associated JVM had seen research split depending on the target of Java code execution, whether on the client or server VMs. The client VM just-in-time compiler (C1) offers fast compilation and a small overall footprint, with the server just-in-time compiler (C2) offering high-performing optimised code at the expense of needing a significantly high proportion of resources. Irrespective of the target, many optimisations have been integrated into the JVM just-in-time compilers over the last three decades. These can be categorised based on compiler tactics, for example, as presented by Ivanov et al. in \cite{Ivanov}: 

\begin{itemize}
    \item Delayed compilation: tiered compilation, or on-stack replacement strategies; 
    \item Proof based techniques: exact type inference, memory value inference, constant folding, null-check elimination, algebraic simplification, common sub-expression elimination; 
    \item Flow sensitive rewrites: conditional constant propagation, flow-carried type narrowing, dead code elimination; 
    \item Language specific techniques: class hierarchy analysis, devirtualisation, symbolic constant propagation, autobox elimination, escape analysis, lock elision and fusion;
    \item Speculative profile based techniques: optimistic nullness and type assertions, optimistic type and array length strengthening, untaken branch pruning, optimistic N-morphic inlining, branch frequency prediction, call frequency prediction; memory and placement transformation: expression hoisting, expression sinking, redundant store elimination, adjacent store fusion; 
    \item Loop transformations: loop unrolling, loop peeling, safepoint elimination, iteration range splitting, range check elimination, loop vectorisation; 
    \item Global code shaping: inlining (graph integration), global code motion, heat-based code layout, switch balancing, throw inlining;
    \item Control flow graph transformation: local code scheduling, local code bundling, delay slot filling, graph-coloring register allocation, linear scan register allocation, live range splitting, copy coalescing, constant splitting, copy removal, address mode matching, instruction peep-holing, and DFA-based code generation.
\end{itemize}

The literature on evaluations of the performance of interpretive execution relative to JIT execution is somewhat sparse regarding contemporary work. The most recent work with any reference is provided by Magalhaes et al. \cite{Magalhaes}. Their work reports that JVM interpretive execution is up to 24 times slower than JIT execution. Their observations relate to the workloads composing the NAS Parallel Benchmarks suite \cite{NAS}. Earlier literature, and in particular the work of Wentworth et al. \cite{Wentworth}, reported as part of their results the differences in JVM interpreter performance relative to Java JIT execution. Although their results are associated with the traditional sorting algorithms: selection sort, bubble sort, recursive quick sort, and heap sort, they showed interpretive execution to be between 6 and 9 times slower than JIT execution. Other early work analysed the performance of several JVM implementations, those integrating a JIT compiler, those with a direct compiler, bytecode translators, and bytecode processor against JVM interpretive execution, for example, Kazi et al. in \cite{Kazi}. The results associated with the JIT compilation showed performance benefits in the range of approximately 2.5 times to 6.6 times over interpretation. The direct compiler implementations showed the most significant performance gains, such as the NET, Caffeine, and HPCJ compilers, with gains 10 and 20 times faster than interpretation.

In relation to identifying differences in performance across JVM variants, other early research analysed the performance of various JVM implementations, including those integrating JIT compilers, direct compilers, bytecode translators, and bytecode processors, in comparison to JVM interpretive execution. For example, Kazi et al. \cite{Kazi} found that JIT compilation resulted in performance gains ranging from approximately 2.5 to 6.6 times over interpretation. Direct compiler implementations, such as the NET, Caffeine, and HPCJ compilers, showed the best performance gains, with improvements ranging from 10 to 20 times faster than interpretation. The most recent literature reporting on the performance of JVM variants would seem to be that detailed by Lambert et al. \cite{Lambert2022}. Although their work focuses solely on the template interpreter, they provide a detailed review across six OpenJDK versions built with one of four GNU GCC compilers. Their work identifies significant effect differences in template interpreter performance depending on workload type and JVM version build toolchain compiler versions.

\section{Method}\label{sec3}
In this section, we present the methodology followed in this study. We first present an overview of the hardware and operating system, followed by an overview of the OpenJDK JRE versions, followed by an introduction to the benchmark workloads.

\subsection{Hardware and Operating System}
\begin{figure}[ht] 
\centering
\begin{tabular}{c}
\includegraphics[width=45mm]{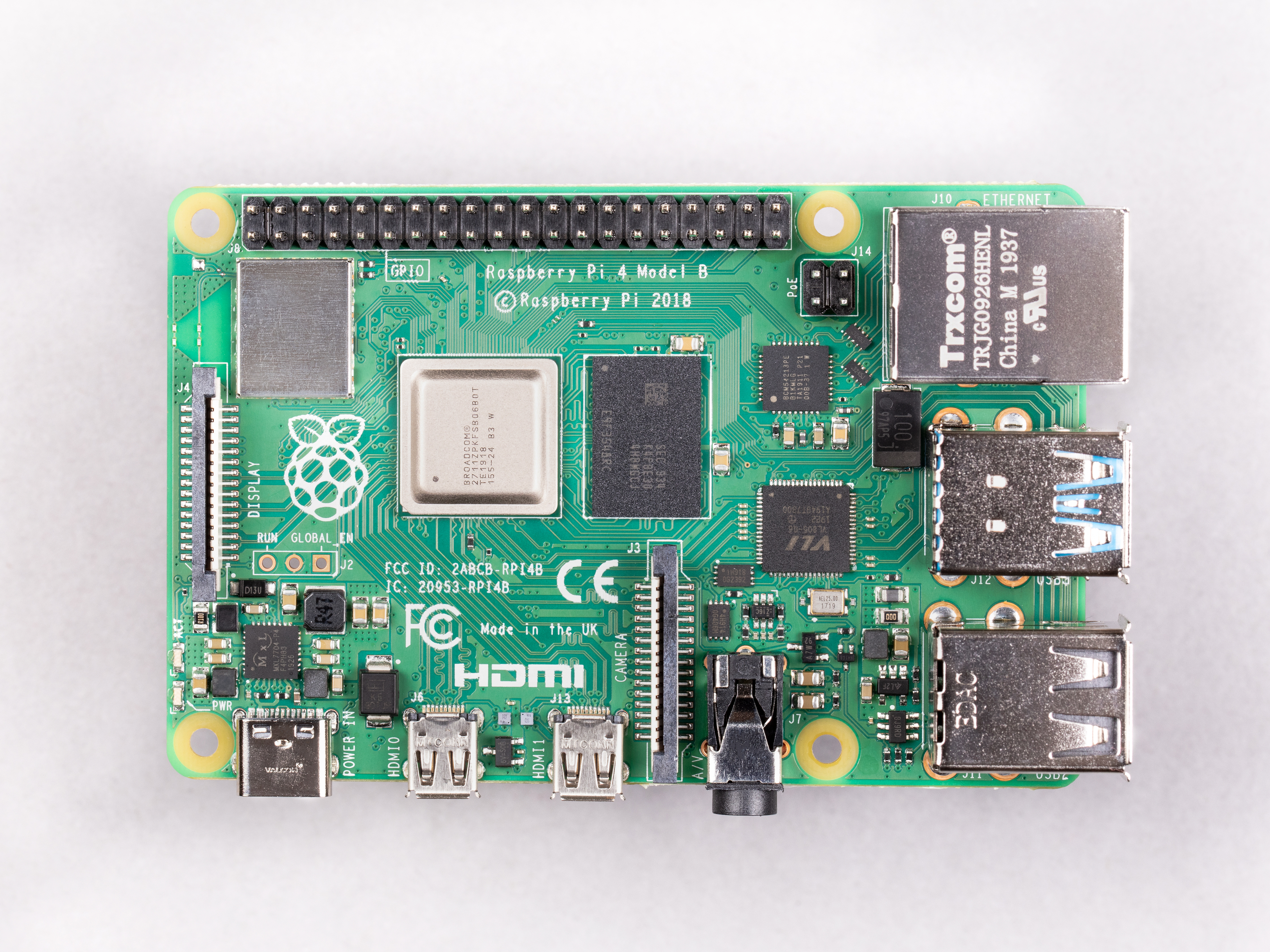} \\
\end{tabular}
\caption{Raspberry Pi4 Model B board peripheral device layout. Detailing position of Broadcom BCM2711 processor, RAM, USB Type-C power supply, and~network ports. Source:~\cite{RaspberrySpec}.}
\label{fig:Hardware}
\end{figure}

Our experimental set-up consists of a cluster of 10 Raspberry Pi4 Model B development boards, integrated with Broadcom BCM2711, Quad-core Cortex-A72 (ARM v8) 64-bit system-on-chip processors with an internal main reference clock speed of 1.5 GHz. Each Raspberry Pi has a 32 KB L1 data cache, a 48 KB L1 instruction cache per core, a 1 MB L2 cache, and 4 GB of LPDDR4-3200 SDRAM. An IEEE 802.11ac wireless interface provides 2.4 GHz and 5.0 GHz network connectivity, in addition to Bluetooth 5.0, 2 USB 3.0 ports and 2 USB 2.0 ports. All Raspberry Pi's were powered from mains. The maximum tolerable current is 3 Amperes \cite{RaspberryBCM2711}. We present a specification diagram in Figure \ref{fig:Hardware}. 

We ran each Raspberry Pi4 Model B with the Raspbian Buster Light Debian distribution of Linux, Kernel version 4.19. We loaded the operating system (OS) through an embedded Micro-SD card slot. All Raspberry Pis had a clone of the same OS installed. The Raspbian Buster Light distribution is a non-windowing version providing only a command-line interface. All Raspberry Pi4 were SSH-enabled. 

\subsection{OpenJDK JREs and OpenJDK JVMs}
We assessed the performance differences between tiered execution and interpretive execution on 100 variants of OpenJDK JRE JVMs. In particular, our assessment is relative to six versions of OpenJDK, namely, versions 9, 10, 11, 12, 13, and 14. We built each OpenJDK JRE with one of four GNU GCC compiler versions, namely versions 5.5.0, 6.5.0, 7.3.0, and 8.3.0. All OpenJDK JRE JVMs were executed with one of five garbage collector variants specified for heap memory management, namely, the Garbage-First Garbage Collector (G1GC), the Serial Garbage Collector (SGC), the Parallel Garbage Collector (PGC), the Concurrent Mark and Sweep Collector (CGC), and the variant of the Concurrent Mark and Sweep Collector with a parallel collection (CPGC) of the new object heap. The CPGC collector was only available in OpenJDK JRE version 9, and was removed as an option in later versions. In total, 100 OpenJDK JRE JVMs were evaluated and assessed in interpretive and tiered modes of execution. All JVMs were initialised with a minimum heap size of 2GB. We indicated interpretive execution mode by passing the \texttt{-Xint} JVM parameter. 

\begin{table}[hb] 
\centering
\small
\caption{{The Renaissance Benchmark Suite, which comprises 24 applications, has been categorised into different application categories. Below is a list of these categories along with a brief description of each application. This information has been sourced from~\cite{Renaissance}.}.  \label{tab:RenaissanceApplications}} 
\begin{tabular}{lll}
\toprule
\textbf{Category} & \textbf{Application} & \textbf{Description} \\
\midrule
Apache-spark    & \texttt{chi-square}        & Chi-square test from Spark MLlib. \\
                & \texttt{dec-tree}          & Random Forest algorithm from the Spark ML library. \\
                & \texttt{gauss-mix}         & Gaussian mixture model using expectation-maximization. \\
                & \texttt{log-regression}    & Logistic Regression algorithm from the Spark ML library. \\
                & \texttt{movie-lens}        & Recommends movies using the ALS algorithm. \\
                & \texttt{naive-bayes}       & Multinomial Naive Bayes algorithm from the Spark ML library. \\
                & \texttt{page-rank}         & PageRank iterations, using RDDs. \\ 
\midrule
Concurrency     & \texttt{akka-uct}          & Unbalanced Cobwebbed Tree actor workload in Akka. \\
                & \texttt{fj-kmeans}         & K-means algorithm using the fork/join framework. \\
                & \texttt{reactors}          & Benchmarks inspired by the Savina microbenchmark workloads \\
\midrule
Database        & \texttt{db-shootout}       & Shootout test using several in-memory databases. \\
                & \texttt{neo4j-analytics}   & Neo4J graph queries against a movie database. \\
\midrule
Functional      & \texttt{future-genetic}    & Genetic algorithm using the Jenetics library and futures. \\
                & \texttt{mnemonics}          & Solves the phone mnemonics problem using JDK streams. \\
                & \texttt{par-mnemonics}     & Solves the phone mnemonics problem using parallel JDK streams. \\
                & \texttt{rx-scrabble}       & Solves the Scrabble puzzle using the Rx streams. \\
                & \texttt{scrabble}          & Solves the Scrabble puzzle using JDK Streams. \\
\midrule
Scala           & \texttt{dotty}             & Dotty compiler on a set of source code files. \\
                & \texttt{philosophers}      & Solves a variant of the dining philosophers problem. \\
                & \texttt{scala-doku}        & Sudoku Puzzles using Scala collections. \\
                & \texttt{scala-kmeans}      & K-Means algorithm using Scala collections. \\
                & scala-stm  & stmbench7 benchmark using ScalaSTM. \\
\midrule
Web             & \texttt{finagle-chirper}   & Microblogging service using Twitter Finagle. \\
                & \texttt{finagle-http}      & Finagle HTTP requests to a Finagle HTTP server.\\
\bottomrule
\end{tabular}
\end{table}

\subsection{JRE Benchmark Workloads}
We evaluated the performance of all JREs and their associated JVM interpreters and the tiered execution runtimes using the Renaissance Benchmark Suite version 11 \cite{Renaissance}, which consists of 24 workloads categorised into six application classes. These classes include applications designed for execution on the Apache Spark framework, those implementing concurrent programming, database applications, functional paradigm applications compiled to Java bytecode, web applications serving HTTP requests, and applications written in Scala and compiled to Java bytecode. For more information about each application workload, please refer to Table~\ref{tab:RenaissanceApplications}. 

We executed all Renaissance benchmark workloads 30 times, and for both modes of execution, each execution ran on new initialised JVM instances, allowing for the construction of all relevant execution time distributions. The effect of running each benchmark on a new JVM instance ensured that the observations associated with tiered execution were before steady-state behaviour and can be to some degree assumed to capture the behaviour of JVM warmup. The consequence is that the performance difference ratios represent the worst case for tiered execution.

We observed several failed workload runs, specifically for the Renaissance Benchmark application \texttt{db-shootout} that could not run on all JVM versions after OpenJDK11. In addition, several other applications reported deprecated API features, particularly that their dependencies cannot be guaranteed to be supported in later JVM versions, specifically after OpenJDK9. None of the deprecated API features had been removed from the assessed OpenJDK~versions.

\section{Results}\label{sec4}
In this section, we present the results of this study. We first give a descriptive overview of the performance efficiency of tiered execution relative to interpretive execution. We then consider performance efficiency relative to each JRE workload. Next, we consider effect differences based on the JRE version and effect differences based on the GNU GCC compiler version used within the JRE build toolchain. We then present the results associated with the effects of changing garbage collection algorithm on the performance difference between interpreted relative to tiered execution. Finally, we present our results associated with several linear regression models, regressing workload category, and several workload metrics that capture constructs such as concurrency and parallelism execution, as well as the typical metrics associated with object-oriented behaviour.

\subsection{Overall differences in performance between interpretive execution and tiered execution}
\begin{figure}[ht]
    \small
    \centering
    \begin{tabular}{cc}
    \includegraphics[width=40mm, height=40mm]{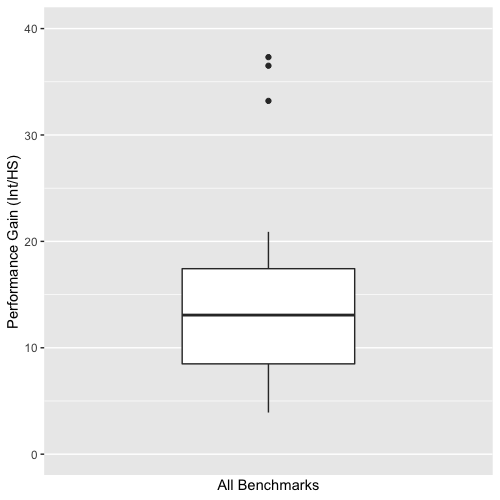} &
    \includegraphics[width=80mm, height=40mm]{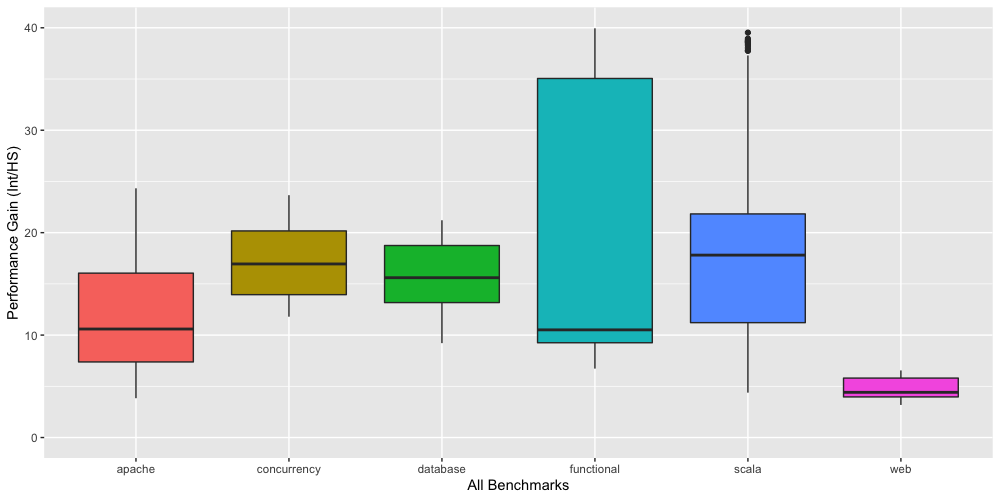}
    \\
    (a) & (b) \\
    \end{tabular}
    \caption{A box-and-whisker plot showing OpenJDK JRE JVM performance efficiency associated with tiered execution relative to interpretive execution, irrespective of workload (a) and the Renaissance workload categories (b).}
    \label{fig:Boxplot-INTvHS-AllBench}
\end{figure}

Irrespective of workload type, tiered execution is significantly more efficient than interpretive execution, with a performance gap ranging from 4 to 37 times more efficiency. On average, tiered execution is approximately 15 times more efficient, with a standard deviation of approximately 9 times. Median efficiency is recorded at 13 times in favour of tiered execution, with associated first and third-quartile efficiency recorded at 8 and 17 times. Figure \ref{fig:Boxplot-INTvHS-AllBench} (a) presents a box-and-whisker plot depicting the distribution of efficiency gains. 

Categorising the performance gain measurements based on the six workload categories defined within the Renaissance Benchmark Suite, Figure \ref{fig:Boxplot-INTvHS-AllBench} (b) shows their respective distributions. The performance ratio that indicates the smallest gain for tiered execution relative to interpretive execution is associated with the two web application workloads: \texttt{finagle-chirper} and \texttt{finagle-http}. The JRE exhibits the widest variation in performance while under load from the five functional application workloads. An inspection of median performance, based on workload category, shows statistically significant differences in category medians, $\chi^{2} = 675.43$, $df = 5$, $p < .001$. With the exception of three pairwise comparison cases, namely: \texttt{scala} v \texttt{concurrency}, \texttt{scala} v \texttt{database}, and \texttt{scala} v \texttt{functional}, the results of the remaining twelve pairwise Mann-Whitney U tests showed statistically significant differences in median performance ratios (all $p < .001$).

\begin{figure}[ht]
\small
\centering
\begin{tabular}{cc}
\includegraphics[width=130mm, height=45mm]{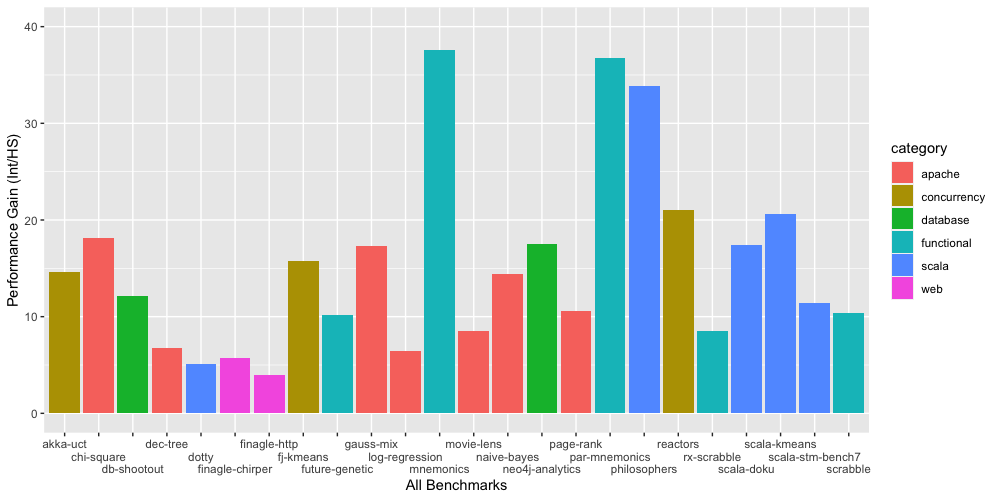} \\
\end{tabular}
\caption{A bar chart plot depicting OpenJDK JRE JVM performance efficiency associated with tiered execution relative to interpretive execution, detailing the average OpenJDK JRE performance efficiency under load from each of the 24 Renaissance benchmark workloads.}
\label{fig:Hist-INTvHS-EachBench}
\end{figure}

Examining performance difference ratios at the benchmark level, we present the relative performance efficiency of tiered execution over interpretive execution for each benchmark workload in Figure \ref{fig:Hist-INTvHS-EachBench}. The horizontal axis lists the 24 Renaissance benchmark workloads, and the vertical axis lists average performance efficiency as the ratio of interpretive execution runtime to tiered execution runtime. Individual benchmark bars are coloured relative to the category of application the benchmark belongs to. In all cases, tiered execution outperformed interpretive execution. For the three application workloads: \texttt{mnemonics}, \texttt{par-mnemonics}, and \texttt{philosophers}, JRE tiered execution performance efficiency was in excess of 30 times better than interpretive execution. Ten application workloads had tiered execution performance efficiency between 10 times and 20 times better than interpretive execution. The JRE JVM running in tiered execution mode showed performance gains between 3 and 10 times for the remaining application workloads. 

Table \ref{tb:Desc-INTvHS-EachBench} presents a more detailed listing of the performance differences for each Renaissance benchmark workload. For example, OpenJDK JRE JVM performance differences between tiered and interpretive execution, for the \texttt{akka-uct} workload, are on average 14.26 times more efficient with an associated standard deviation of 2.20.   

\begin{table}[ht]
  \centering
  \small
  \caption{Detailed listing of exact performance gain observations for each of the 24 Renaissance workloads. Listing the respective benchmark workload names, Benchmark, the average performance difference ratio across all OpenJDK runs (M), and the associated standard deviation (SD).}
  \label{tb:Desc-INTvHS-EachBench}
  \begin{tabular}{lrrlrrlrr}
  \toprule
  \textbf{Benchmark} & \textbf{M} & \textbf{SD} & 
  \textbf{Benchmark} & \textbf{M} & \textbf{SD} &
  \textbf{Benchmark} & \textbf{M} & \textbf{SD} \\
  \midrule
akka-uct        & 14.66 & 2.20 & future-genetic  & 10.18 & 0.59 & par-mnemonics    & 36.73 & 2.45 \\
chi-square      & 18.14 & 2.11 & gauss-mix       & 17.26 & 2.19 & philosophers     & 33.85 & 4.07 \\
db-shootout     & 12.14 & 1.48 & log-regression  & 6.48  & 0.83 & reactors         & 21.00 & 1.16 \\
dec-tree        & 6.79  & 0.72 & mnemonics       & 37.62 & 2.90 & rx-scrabble      & 8.56  & 0.51 \\
dotty           & 5.14  & 0.29 & movie-lens      & 8.48  & 0.92 & scala-doku       & 17.40 & 1.63 \\
finagle-chirper & 5.75  & 0.42 & naive-bayes     & 14.40 & 2.05 & scala-kmeans     & 20.61 & 1.93 \\
finagle-http    & 3.92  & 0.25 & neo4j-analytics & 17.53 & 2.34 & scala-stm-bench7 & 11.45 & 0.76 \\
fj-kmeans       & 15.75 & 1.73 & page-rank       & 10.62 & 1.30 & scrabble         & 10.35 & 0.77 \\
  \bottomrule
  \end{tabular}
\end{table}

\begin{figure}[ht]
\small
\centering
\begin{tabular}{c}
\includegraphics[width=130mm, height=45mm]{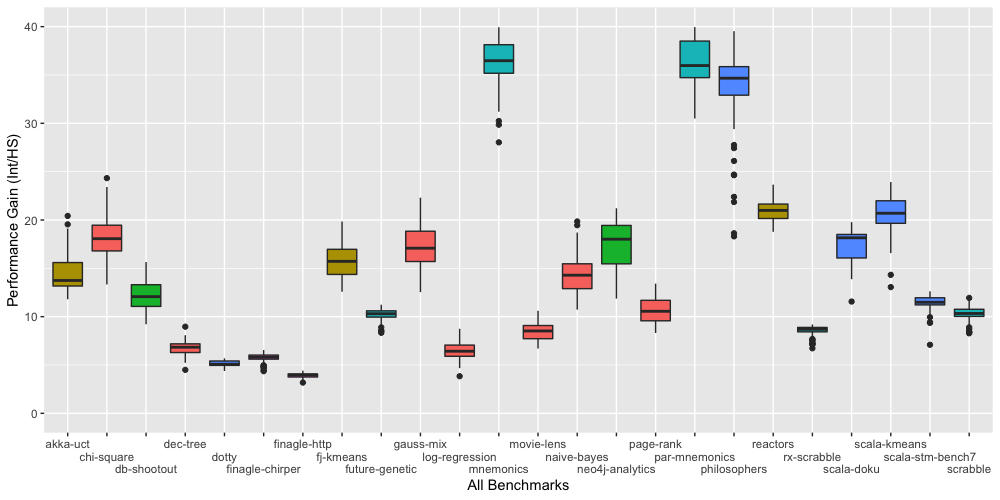} \\
\end{tabular}
\caption{Box-and-whisker plots depicting OpenJDK JRE JVM performance efficiency associated with tiered execution relative to interpretive execution, detailing performance efficiency of the OpenJDK JRE under load from each of the 24 Renaissance benchmark workloads.}
\label{fig:Boxplot-INTvHS-EachBench}
\end{figure}

In Figure \ref{fig:Boxplot-INTvHS-EachBench}, we present a more detailed overview of the performance efficiency associated with tiered execution relative to interpretive execution for each of the 24 Renaissance workloads. Each relevant box-and-whisker plot is coloured based on the applications category defined by the Renaissance Benchmark Suite. Focusing on the widths of the associated box-and-whisker plots shows a collection of workloads with minimal variation, for example, \texttt{dec-tree}, \texttt{dotty}, \texttt{finagle-chirper}, \texttt{finagle-http}, \texttt{future genetic}, \texttt{rx-scrabble}, and \texttt{scrabble}.

\subsection{Differences in performance between interpretive execution and tiered execution across OpenJDK JRE version}
This section presents the results associated with analysing performance difference ratios for interpretive execution relative to tiered execution at the OpenJDK JRE version level. In particular, we detail the results related to an analysis of six OpenJDK JRE versions, namely: versions 9 through 14. In addition, we consider the performance difference effects associated with the six workload categories assessed within each OpenJDK JRE version.

\begin{figure}[ht]
    \small
    \centering
    \begin{tabular}{c}
    \includegraphics[width=130mm, height=45mm]{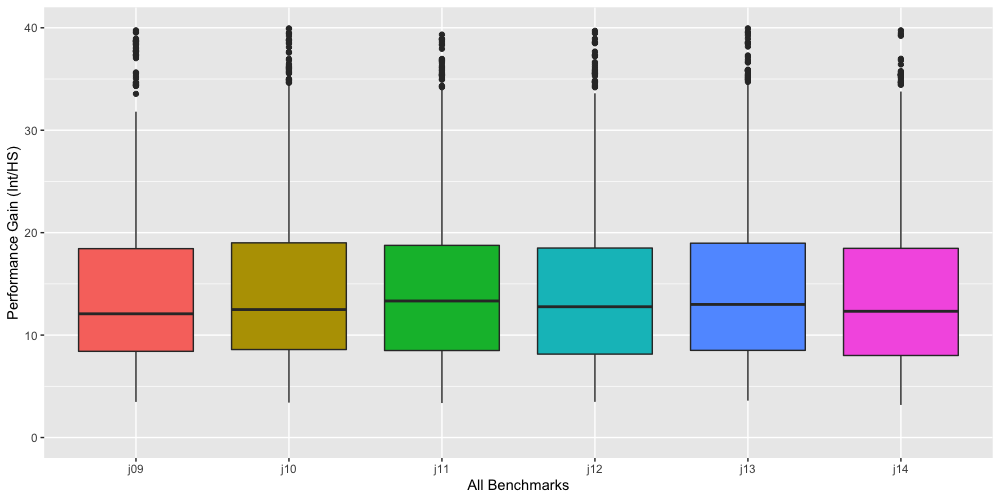} \\
    \end{tabular}
    \caption{Box-and-whisker plots depicting OpenJDK JRE version performance efficiency associated with tiered execution relative to interpretive execution.}
    \label{fig:Boxplot-INTvHS-EachJVM-AllBench} 
\end{figure}

In Figure \ref{fig:Boxplot-INTvHS-EachJVM-AllBench}, we present several box-and-whisker plots depicting the performance efficiency distributions associated with each of the six OpenJDK JRE versions analysed. All distributions show similar characteristics in that irrespective of the OpenJDK JRE version, 75\% of performance efficiency measures are less than 20. OpenJDK JRE versions 9 and 10 have the smallest median performance efficiency relative to the other four OpenJDK JRE versions. We undertook a Kruskal Wallace test for differences in distribution medians, the results of the Kruskal Wallace H-test indicating that there was no significant difference in median OpenJDK JRE performance gap based on the version of OpenJDK JRE analysed, $\chi^2 = 1.07$, $df = 5$, $p = .957$.

\begin{figure}[ht]
    \small
    \centering
    \begin{tabular}{c}
    \includegraphics[width=130mm, height=45mm]{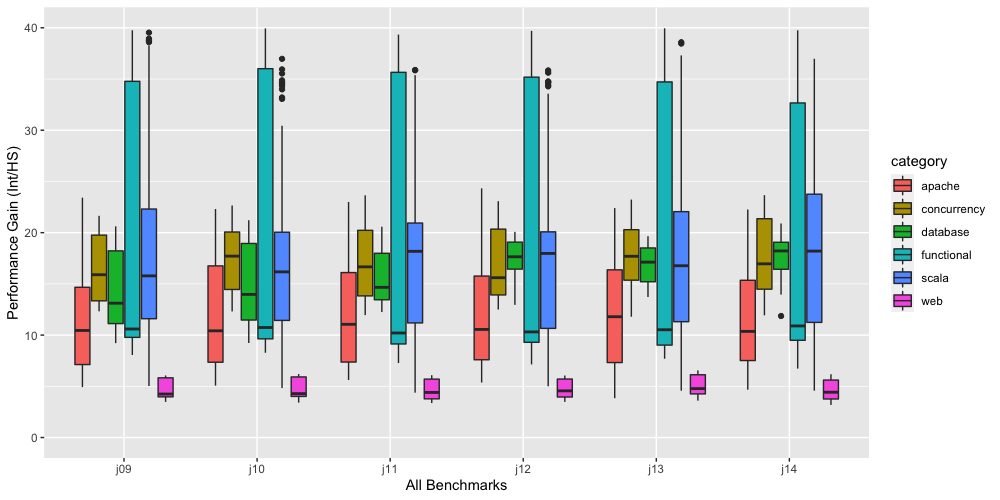} \\
    \end{tabular}
    \caption{Box-and-whisker plots depicting OpenJDK JRE version performance efficiency associated with tiered execution relative to interpretive execution across OpenJDK JRE version, detailed at the workload category level.}
    \label{fig:Boxplot-INTvHS-EachJVM-EachCategory} 
\end{figure}

We present a finer-grained assessment of the performance gap between interpretive and tiered execution based on workload category in Figure \ref{fig:Boxplot-INTvHS-EachJVM-EachCategory}. The horizontal axis lists the six OpenJDK JRE versions, with a categorisation of workload type presented within each version. 

We assessed if median performance ratios differed depending on the workload category for each OpenJDK JRE version. The results of Kruskal-Wallace H-tests indicated statistically significant differences in median performance ratios between workload categories for each OpenJDK JRE version (all $p < .05$). Across all six OpenJDK JRE versions, similar results were observed concerning assessing differences in pairwise median performance ratios. In general, we observed no statistically significant differences between pairwise comparisons of the \texttt{concurrency}, \texttt{functional}, and \texttt{scala} workload types, irrespective of OpenJDK JRE version (all $p > .05$). In general, the \texttt{apache} and \texttt{web} application workloads had statistically significantly different median performance ratios (all $p < .05$). In addition, the \texttt{apache} and \texttt{web} workload categories were statistically significantly different to each of the other four workload categories (all $p < .05$). 

\begin{figure}[ht]
\footnotesize
\centering
\begin{tabular}{c}
\includegraphics[width=130mm, height=45mm]{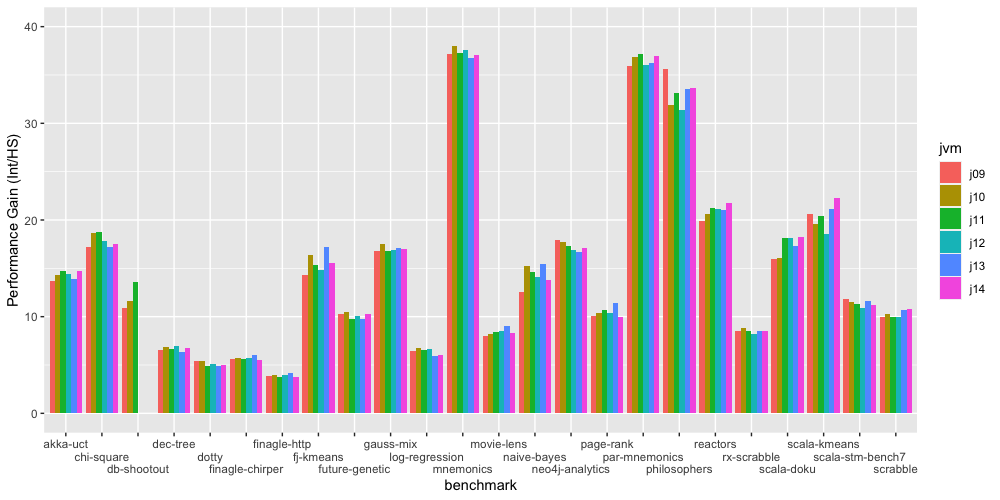} \\
\end{tabular}
\caption{Bar chart depicting OpenJDK JRE version performance efficiency associated with tiered execution relative to interpretive execution detailed across each of the 24 Renaissance benchmark workloads.}
\label{fig:Hist-INTvHS-JVM-EachBench}
\end{figure}

The results suggest no differences across the OpenJDK JRE versions for each workload type. Comparisons indicate variations within each category are similar across all OpenJDK JRE versions. However, the only differences in median performance ratios were associated with the database category of workloads. Specifically, the results of a Kruskal Wallace H-test showed statistically significant differences in performance for database workloads depending on the OpenJDK JRE version used for their execution, $\chi^2 = 14.073$, $df = 5$, $p = .015$. As shown in Figure \ref{fig:Boxplot-INTvHS-EachJVM-EachCategory}, median performance ratios associated with the database category of workload increase, progressing from older to newer versions of OpenJDK JRE. We undertook a Jonckheere-Terpstra test to assess for increasing alternatives, and increases in median performance ratios were confirmed, $JT = 5334$, $p < .001$. Compared across OpenJDK JRE versions, we observed no statistically significant differences for the other six workload categories.

\begin{table}[ht]
  \centering
  \small
  \caption{Detailed listing of exact performance gain observations for each of the 24 Renaissance workloads. The \textbf{Gain} column representing the ratio of interpreter execution runtime to tiered execution runtime.}
  \label{tab:Desc-INTvHS-JVMa-EachBench}
  \begin{tabular}{llr|llr|llr}
  \toprule
  \textbf{JRE} & \textbf{Benchmark} & \textbf{Gain} & \textbf{JRE} & \textbf{Benchmark} & \textbf{Gain} & \textbf{JRE} & \textbf{Benchmark} & \textbf{Gain}  \\
  \midrule
j09	&	akka-uct	&	13.70	&	j09	&	dec-tree	&	6.54	&	j09	&	finagle-http	&	3.86	\\
j10	&	akka-uct	&	14.32	&	j10	&	dec-tree	&	6.90	&	j10	&	finagle-http	&	3.93	\\
j11	&	akka-uct	&	14.68	&	j11	&	dec-tree	&	6.67	&	j11	&	finagle-http	&	3.80	\\
j12	&	akka-uct	&	14.39	&	j12	&	dec-tree	&	6.99	&	j12	&	finagle-http	&	3.93	\\
j13	&	akka-uct	&	13.84	&	j13	&	dec-tree	&	6.38	&	j13	&	finagle-http	&	4.13	\\
j14	&	akka-uct	&	14.70	&	j14	&	dec-tree	&	6.74	&	j14	&	finagle-http	&	3.75	\\
\midrule
j09	&	chi-square	&	17.17	&	j09	&	dotty	&	5.46	&	j09	&	fj-kmeans	&	14.32	\\
j10	&	chi-square	&	18.66	&	j10	&	dotty	&	5.42	&	j10	&	fj-kmeans	&	16.40	\\
j11	&	chi-square	&	18.77	&	j11	&	dotty	&	4.92	&	j11	&	fj-kmeans	&	15.33	\\
j12	&	chi-square	&	17.79	&	j12	&	dotty	&	5.12	&	j12	&	fj-kmeans	&	14.79	\\
j13	&	chi-square	&	17.23	&	j13	&	dotty	&	4.89	&	j13	&	fj-kmeans	&	17.20	\\
j14	&	chi-square	&	17.52	&	j14	&	dotty	&	4.99	&	j14	&	fj-kmeans	&	15.57	\\
\midrule
j09	&	db-shootout	&	10.85	&	j09	&	finagle-chirper	&	5.57	&	j09	&	future-genetic	&	10.24	\\
j10	&	db-shootout	&	11.62	&	j10	&	finagle-chirper	&	5.71	&	j10	&	future-genetic	&	10.48	\\
j11	&	db-shootout	&	13.54	&	j11	&	finagle-chirper	&	5.62	&	j11	&	future-genetic	&	9.80	\\
j12	&	db-shootout	&		&	j12	&	finagle-chirper	&	5.71	&	j12	&	future-genetic	&	10.09	\\
j13	&	db-shootout	&		&	j13	&	finagle-chirper	&	6.02	&	j13	&	future-genetic	&	9.78	\\
j14	&	db-shootout	&		&	j14	&	finagle-chirper	&	5.54	&	j14	&	future-genetic	&	10.23	\\
\midrule
j09	&	gauss-mix	&	16.83	&	j09	&	movie-lens	&	7.95	&	j09	&	page-rank	&	10.08	\\
j10	&	gauss-mix	&	17.56	&	j10	&	movie-lens	&	8.23	&	j10	&	page-rank	&	10.42	\\
j11	&	gauss-mix	&	16.77	&	j11	&	movie-lens	&	8.43	&	j11	&	page-rank	&	10.67	\\
j12	&	gauss-mix	&	16.85	&	j12	&	movie-lens	&	8.49	&	j12	&	page-rank	&	10.40	\\
j13	&	gauss-mix	&	17.13	&	j13	&	movie-lens	&	8.99	&	j13	&	page-rank	&	11.44	\\
j14	&	gauss-mix	&	16.96	&	j14	&	movie-lens	&	8.30	&	j14	&	page-rank	&	9.99	\\
  \bottomrule
  \end{tabular}
\end{table}

Focusing on JRE performance differences based on the JRE version used to execute the workloads, Figure \ref{fig:Hist-INTvHS-JVM-EachBench} depicts those differences for each benchmark workload. Performance differences based on the JRE version are evident, being more pronounced for specific workloads, for example, philosophers, where JRE version 9 provides increased performance efficiency for tiered execution. Across all workloads, JRE performance efficiency can be increased by, on average, 6\% with an associated standard deviation of 4\% through a change in the OpenJDK JRE version. The maximum possible increase is approximately 20\%. 

\begin{table}[ht]
  \centering
  \small
  \caption{Detailed listing of exact performance gain observations for each of the 24 Renaissance workloads. The \textbf{Gain} column representing the ratio of interpreter execution runtime to tiered execution runtime.}
  \label{tab:Desc-INTvHS-JVMb-EachBench}
  \begin{tabular}{llr|llr|llr}
  \toprule
  \textbf{JRE} & \textbf{Benchmark} & \textbf{Gain} & \textbf{JRE} & \textbf{Benchmark} & \textbf{Gain} & \textbf{JRE} & \textbf{Benchmark} & \textbf{Gain}  \\
  \midrule
j09	&	log-regression	&	6.45	&	j09	&	naive-bayes	&	12.59	&	j09	&	par-mnemonics	&	35.98	\\
j10	&	log-regression	&	6.71	&	j10	&	naive-bayes	&	15.20	&	j10	&	par-mnemonics	&	36.87	\\
j11	&	log-regression	&	6.59	&	j11	&	naive-bayes	&	14.66	&	j11	&	par-mnemonics	&	37.13	\\
j12	&	log-regression	&	6.65	&	j12	&	naive-bayes	&	14.09	&	j12	&	par-mnemonics	&	35.99	\\
j13	&	log-regression	&	5.90	&	j13	&	naive-bayes	&	15.47	&	j13	&	par-mnemonics	&	36.26	\\
j14	&	log-regression	&	6.08	&	j14	&	naive-bayes	&	13.80	&	j14	&	par-mnemonics	&	36.96	\\
\midrule
j09	&	mnemonics	&	37.16	&	j09	&	neo4j-analytics	&	17.88	&	j09	&	philosophers	&	35.66	\\
j10	&	mnemonics	&	38.04	&	j10	&	neo4j-analytics	&	17.73	&	j10	&	philosophers	&	31.88	\\
j11	&	mnemonics	&	37.25	&	j11	&	neo4j-analytics	&	17.35	&	j11	&	philosophers	&	33.08	\\
j12	&	mnemonics	&	37.62	&	j12	&	neo4j-analytics	&	16.94	&	j12	&	philosophers	&	31.33	\\
j13	&	mnemonics	&	36.76	&	j13	&	neo4j-analytics	&	16.65	&	j13	&	philosophers	&	33.57	\\
j14	&	mnemonics	&	37.09	&	j14	&	neo4j-analytics	&	17.06	&	j14	&	philosophers	&	33.69	\\
\midrule
j09	&	reactors	&	19.85	&	j09	&	scala-doku	&	15.92	&	j09	&	scala-stm-bench7	&	11.86	\\
j10	&	reactors	&	20.58	&	j10	&	scala-doku	&	16.05	&	j10	&	scala-stm-bench7	&	11.53	\\
j11	&	reactors	&	21.19	&	j11	&	scala-doku	&	18.13	&	j11	&	scala-stm-bench7	&	11.31	\\
j12	&	reactors	&	21.09	&	j12	&	scala-doku	&	18.13	&	j12	&	scala-stm-bench7	&	10.92	\\
j13	&	reactors	&	21.01	&	j13	&	scala-doku	&	17.27	&	j13	&	scala-stm-bench7	&	11.59	\\
j14	&	reactors	&	21.72	&	j14	&	scala-doku	&	18.22	&	j14	&	scala-stm-bench7	&	11.16	\\
\midrule
j09	&	rx-scrabble	&	8.52	&	j09	&	scala-kmeans	&	20.64	&	j09	&	scrabble	&	10.01	\\
j10	&	rx-scrabble	&	8.80	&	j10	&	scala-kmeans	&	19.57	&	j10	&	scrabble	&	10.30	\\
j11	&	rx-scrabble	&	8.48	&	j11	&	scala-kmeans	&	20.44	&	j11	&	scrabble	&	9.93	\\
j12	&	rx-scrabble	&	8.23	&	j12	&	scala-kmeans	&	18.58	&	j12	&	scrabble	&	9.93	\\
j13	&	rx-scrabble	&	8.55	&	j13	&	scala-kmeans	&	21.12	&	j13	&	scrabble	&	10.65	\\
j14	&	rx-scrabble	&	8.48	&	j14	&	scala-kmeans	&	22.32	&	j14	&	scrabble	&	10.79	\\
  \bottomrule
  \end{tabular}
\end{table}

A more detailed listing of the performance efficiency gains associated with tiered execution relative to interpretive execution is listed in Tables \ref{tab:Desc-INTvHS-JVMa-EachBench} and \ref{tab:Desc-INTvHS-JVMb-EachBench}. Both tables list the performance efficiency ratios associated with the 24 Renaissance workloads running on each of the six OpenJDK JRE JVMs.

\subsection{Differences in performance between interpretive execution and tiered execution across OpenJDK JRE build version}

\begin{figure}[ht]
    \tiny
    \centering
    \begin{tabular}{cc}
    \includegraphics[width=50mm, height=45mm]{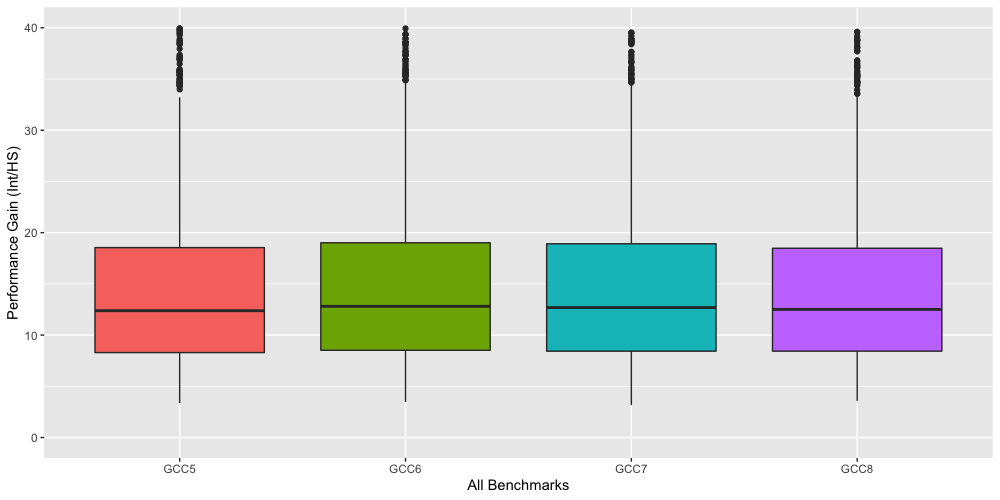}
    &
    \includegraphics[width=80mm, height=45mm]{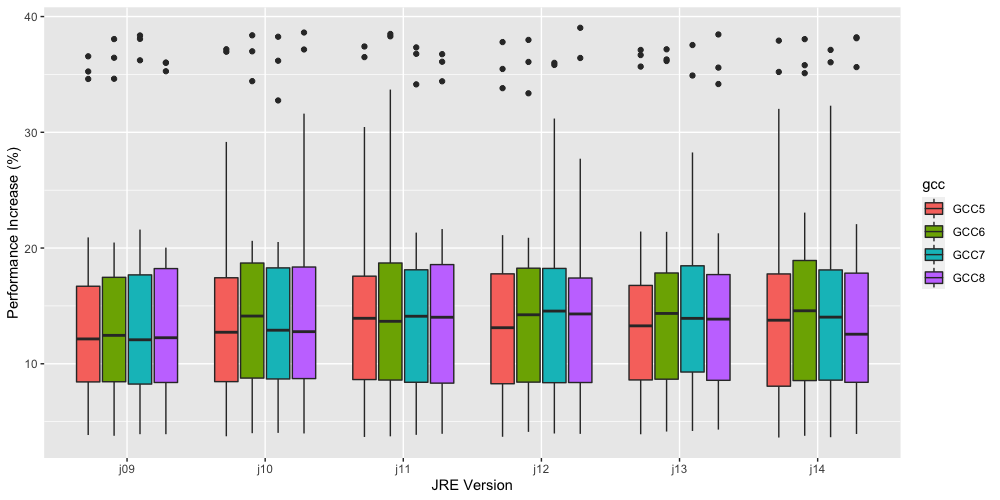} \\
    (a) Pooled Performance Ratios & (b) Across OpenJDK JRE versions \\
    \end{tabular}
    \caption{Box-and-whisker plots depicting OpenJDK JRE version performance ratio for interpretive execution relative to tiered execution detailing those ratios for each GNU GCC compiler version (a), in addition to, GNU GCC compiler build toolchain effects associated with each OpenJDK JRE version (b).}
    \label{fig:Boxplot-INTvHS-JVM-GCC-EachBench} 
\end{figure}

The results from an analysis of the effect of changing the GNU GCC compiler version in the build toolchain of the JRE are presented in Figure \ref{fig:Boxplot-INTvHS-JVM-GCC-EachBench} (a) and (b). Concerning differences in performance ratio based solely on the GNU GCC version used in the build toolchain of the OpenJDK versions (Figure \ref{fig:Boxplot-INTvHS-JVM-GCC-EachBench} (a)), we observed no statistically significant difference, $\chi^2 = 1.088$, $df = 3$, $p = .780$. Similarly, exploring for differences in median performance ratios at the OpenJDK JRE version level (Figure \ref{fig:Boxplot-INTvHS-JVM-GCC-EachBench} (b)),  we observed no statistically significant difference in median performance ratios between GNU GCC versions (all $p > .05$).

\begin{figure}[ht]
    \tiny
    \centering
    \begin{tabular}{cc}
    \includegraphics[width=65mm, height=30mm]{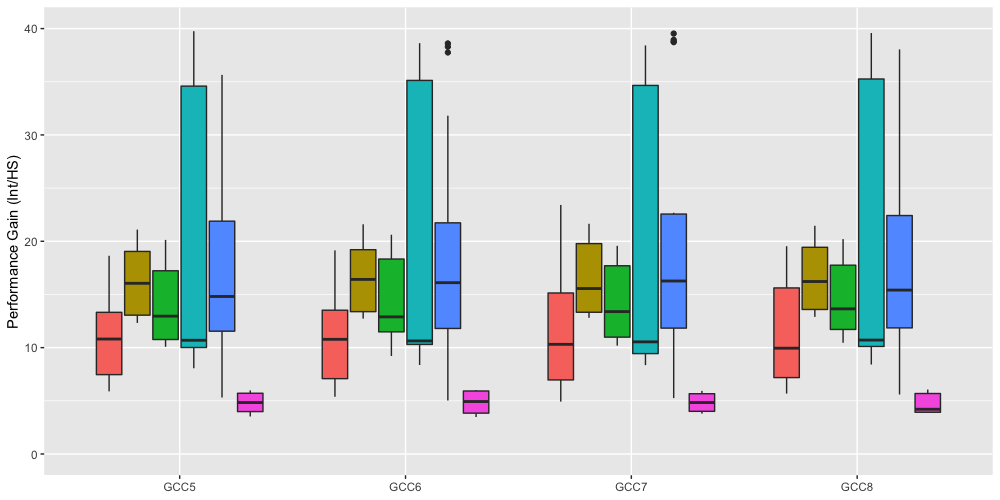} & 
    \includegraphics[width=65mm, height=30mm]{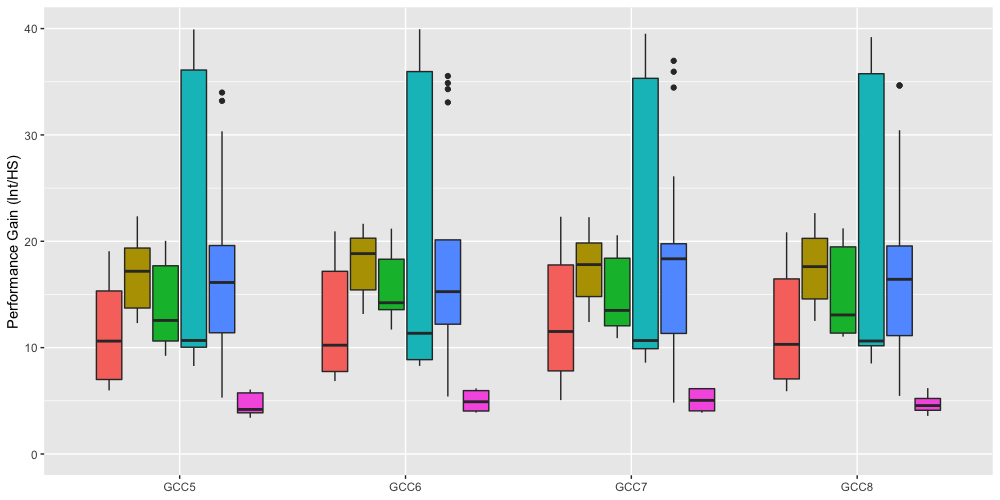}
    \\
    (a) OpenJDK JRE Version 9 & (b) OpenJDK JRE Version 10 \\
    
    \includegraphics[width=65mm, height=30mm]{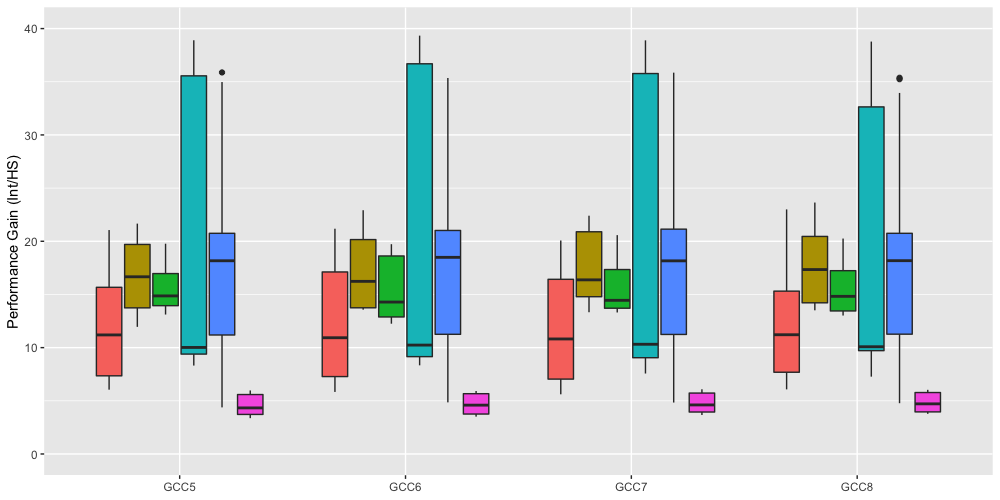} & 
    \includegraphics[width=65mm, height=30mm]{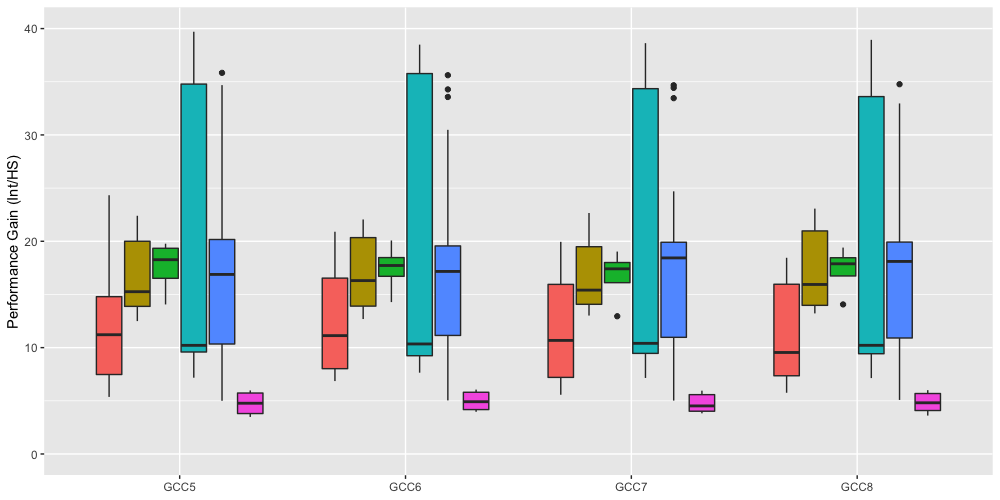}
    \\
    (c) OpenJDK JRE Version 11 & (d) OpenJDK JRE Version 12 \\

    \includegraphics[width=65mm, height=30mm]{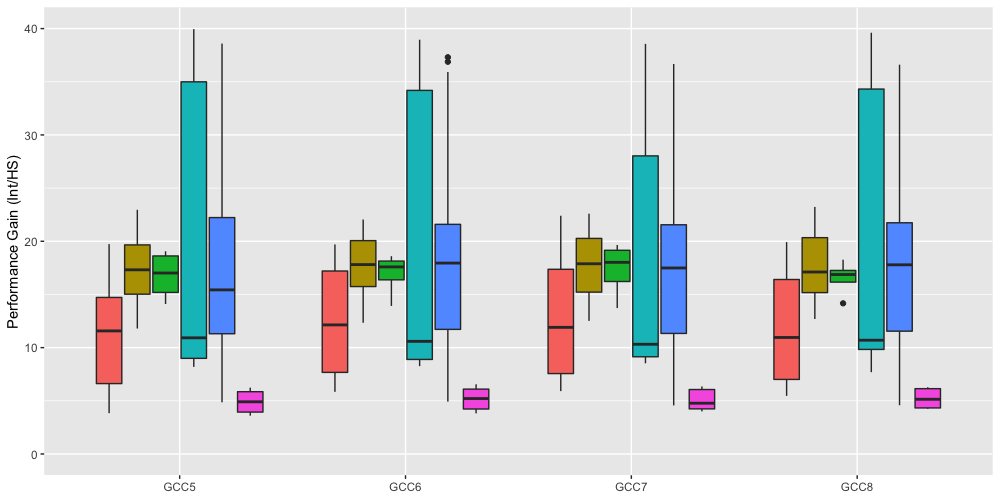} & 
    \includegraphics[width=65mm, height=30mm]{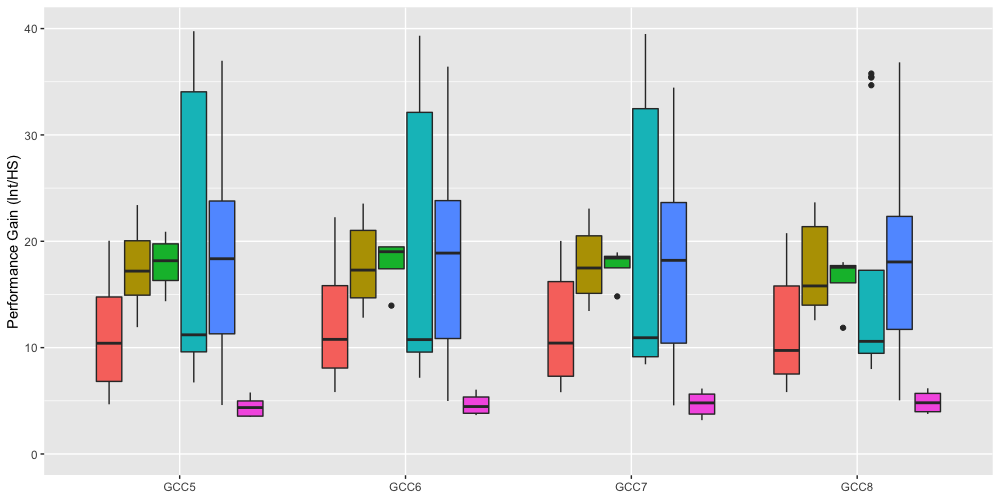}
    \\
    (e) OpenJDK JRE Version 13 & (f) OpenJDK JRE Version 14 \\
    \end{tabular}
    \caption{Box-and-whisker plot panels depicting OpenJDK JRE version performance ratios, detailing each workload category for with each GNU GCC compiler version.}
    \label{fig:Boxplot-INTvHS-EachGCC-EachCategory} 
\end{figure}

In Figure \ref{fig:Boxplot-INTvHS-EachGCC-EachCategory}, we present six panels, one for each OpenJDK JRE version, each panel depicting box-and-whisker plot representations of performance ratio differences distribution for each of the six categories of workload across each of the four GNU GCC versions. An analysis of the differences in median performance showed statistically significant differences based on workload type across all.

\subsection{Differences in performance between interpretive execution and tiered execution based on garbage collection algorithm}
In this section, we present the results associated with analysing the performance difference ratios between interpretive execution and tiered execution, explicitly detailing the effects related to different garbage collector algorithms.

\begin{figure}[ht]
    \tiny
    \centering
    \begin{tabular}{cc}
    \includegraphics[width=50mm, height=45mm]{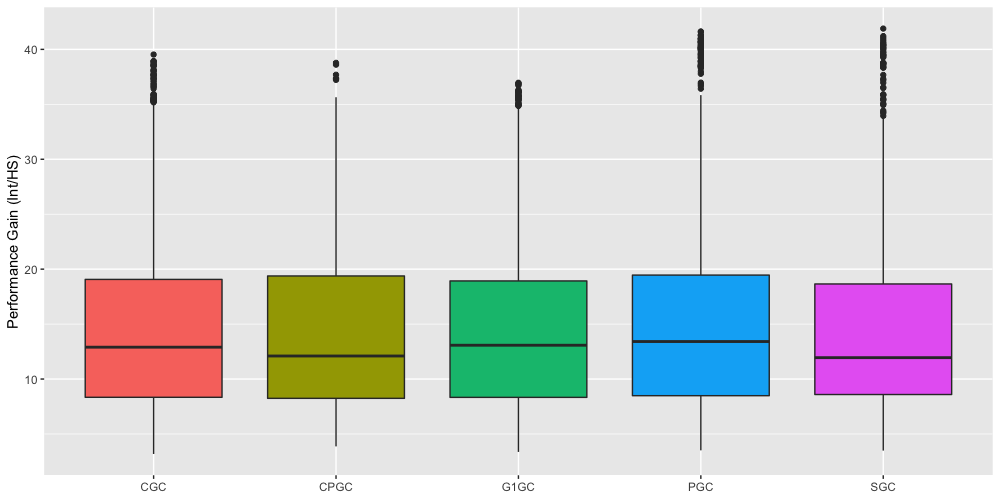}
    &
    \includegraphics[width=80mm, height=45mm]{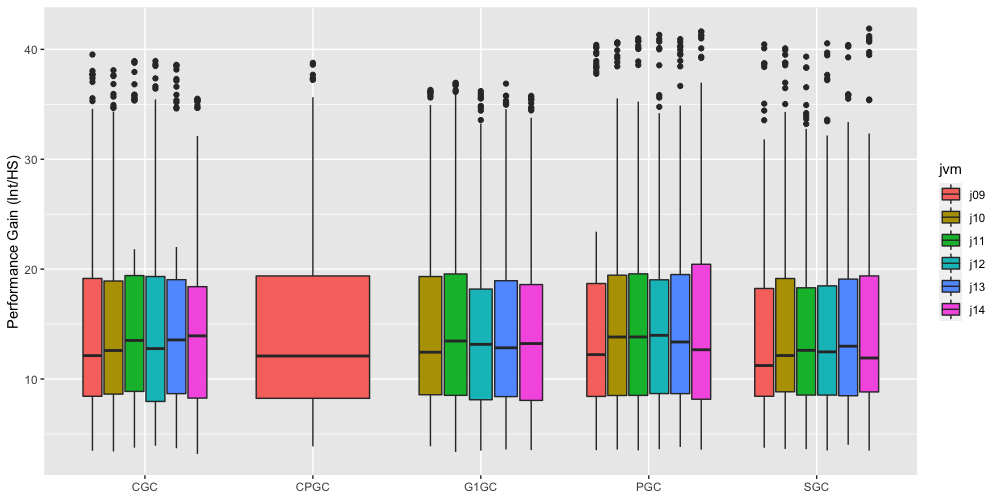} \\
    (a) Pooled Performance Ratios & (b) Across OpenJDK JRE versions \\
    \end{tabular}
    \caption{Box-and-whisker plots depicting OpenJDK JRE version performance ratio for interpretive execution relative to tiered execution detailing those ratios for each garbage collection algorithm assessed (a), in addition to differences in performance across OpenJDK JRE JVM versions while operating with the same heap memory management garbage collector (b).}
    \label{fig:Boxplot-INTvHS-GC-EachJVM} 
\end{figure}

In Figure \ref{fig:Boxplot-INTvHS-GC-EachJVM} (a) we present the pooled performance ratios associated with OpenJDK JRE JVM operation while operating with one of five garbage collection strategies. In particular, with a concurrent mark and sweep algorithm (CGC), concurrent mark and sweep algorithm with a parallel reclamation of the new object heap (CPGC), with the garbage first collector (G1GC), with a parallel collector (PGC), and with the serial garbage collector (SGC). An assessment of pooled differences in median performance ratios between garbage collector algorithms showed no statistically significant differences $\chi^2 = 1.376$, $df = 4$, $p = 0.848$. No evidence of differences in median performance ratios, based on garbage collector, were found for the \texttt{appache}, \texttt{functional}, \texttt{scala}, or \texttt{web} categories of benchmark workloads.

In Figure \ref{fig:Boxplot-INTvHS-GC-EachJVM} (b) we present box-and-whisker plots of the performance ratio distributions associated with each garbage collection algorithm, detailed within each OpenJDK JRE version. For example, concentrating on the concurrent-mark-and-sweep collector (CGC), first group of boxplots in Figure \ref{fig:Boxplot-INTvHS-GC-EachJVM} (b), median performance difference ratios are highest for OpenJDK JRE version 14. The gap between interpretive execution and tiered execution is smallest for OpenJDK JRE version 9. Assessing if there is evidence to suggest differences in median performance ratios holding garbage collection algorithm constant across the OpenJDK JRE versions, we observed no statistically significant difference in median performance difference ratios (all $p > .05)$).

\begin{figure}[ht]
    \tiny
    \centering
    \begin{tabular}{cc}
    \includegraphics[width=65.25mm, height=45mm]{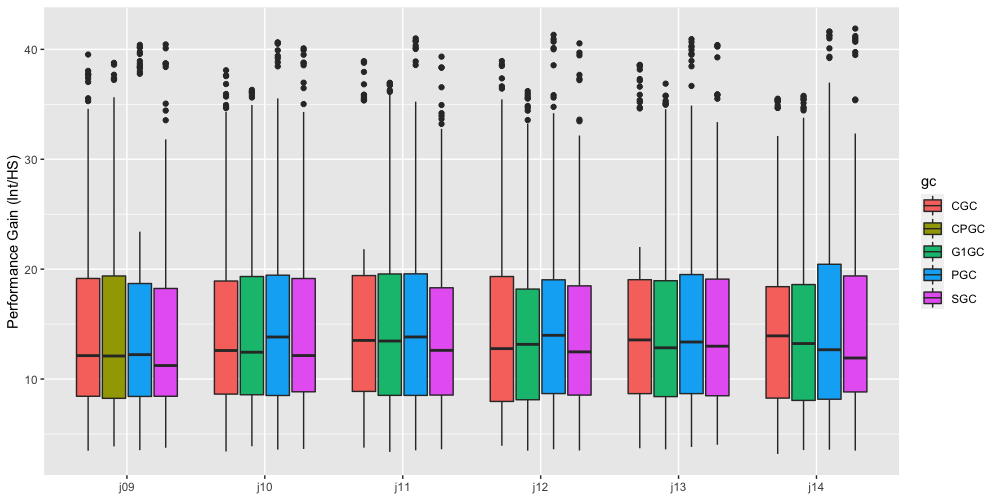} 
    &
     \includegraphics[width=65.25mm, height=45mm]{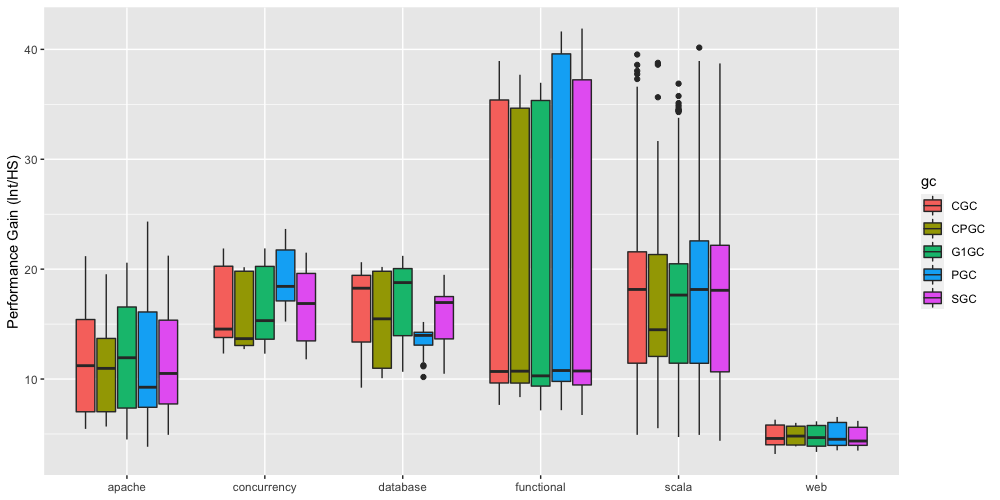} \\
     (a) & (b)
    \end{tabular}
    \caption{Box-and-whisker plots depicting OpenJDK JRE performance ratios for interpretive execution relative to tiered execution, within JRE version for each garbage collection algorithm (a), within each workload category for each garbage collection algorithm (b).}
    \label{fig:Boxplot-INTvHS-EachJVM-EachGC} 
\end{figure}

Assessing if there are differences in performance at the OpenJDK JRE version level, depending on the garbage collection model, Figure \ref{fig:Boxplot-INTvHS-EachJVM-EachGC} (a) presents those results. We found no evidence to suggest differences in median performance ratios between garbage collection algorithms at the OpenJDK JRE version level (all $p > .05)$). Across all OpenJDK JRE versions the serial collector provides the narrowest margin in median performance between interpretive and tiered execution.

We finally partition workloads into one of six workload categories and assess the performance differences of interpreted execution relative to tiered execution. Those results are presented in Figure \ref{fig:Boxplot-INTvHS-EachJVM-EachGC} (b). The horizontal axis lists workload category and within each workload category we depict the performance ratio for each garbage collection algorithm. Of the six workload categories, only the concurrency and database categories of workload had statistically significant differences in median performance ratios based on garbage collection algorithm, $\chi^2 = 41.220$, $df = 4$, $p < .001$; $\chi^2 = 24.862$, $df = 4$, $p < .001$ respectively.

\subsection{Predicting performance difference ratios between interpretive and tiered execution}
This section presents the results associated with constructing several linear models of performance difference ratios. We first detail the effects of regressing the four predictors OpenJDK JRE version, GNU GCC compiler version, Garbage Collection Algorithm, and workload category on performance difference ratios. We then present the results associated with the regression of concurrency-parallelism metrics on performance difference ratios. In particular, the work of Prokopec et al. in \cite{Renaissance} presented a dimensionality analysis of the collection of workloads comprising the Renaissance benchmark suite. They measured and characterised the workloads based on 11 metrics that help capture the nature of the underlying concurrency and parallelism composition, measuring: thread synchronisation calls, waits, notifies, the number of atomic operations, park, CPU, cache, the number of objects, the number of arrays, the number of method invocations, and the number of invoke dynamic calls. 

We assess the predictability of the OpenJDK JRE version, GNU GCC compiler version, Garbage Collection Algorithm, and workload on performance difference ratios between interpretive execution and tiered execution. Hierarchical multiple regression was statistically significant, F (17, 2238) = 40.27; p < .001 and explained 23\% of variance in performance difference ratios. The only statistically significant predictor was the workload category, which was statistically significant at each nominal level. Table \ref{tab:Reg-INTvHS} presents the model coefficient results.

We present the results associated with the regression of the eleven performance metrics on performance ratios in Table \ref{tab:Reg-INTvHS-Categories}. The only statistically significant regression models are regarding the two predictors \texttt{atom} and \texttt{cpu}, accounting for 25\% and 40\% of the variation in performance difference ratios, respectively.

\begin{table}[ht]
  \centering
  \small
  \caption{Hierarchical linear regression model of performance difference ratios between interpretive and tiered execution.}
  \label{tab:Reg-INTvHS}
  \begin{tabular}{lrrrrl}
  \toprule
  \textbf{Coefficient} & \textbf{B} & \textbf{S.E.} & \textbf{t} & \textbf{Pr(>|t|)} & \\
  \midrule   
(Intercept) & 11.19     & 0.70 & 16.00  & .000      & ***    \\
j10         & 0.20      & 0.66 & 0.30   & .766      &        \\
j11         & 0.25      & 0.66 & 0.38   & .704      &        \\
j12         & 0.07      & 0.67 & 0.11   & .911      &        \\
j13         & 0.44      & 0.67 & 0.66   & .512      &        \\
j14         & 0.41      & 0.67 & 0.61   & .543      &        \\
GCC6        & 0.48      & 0.50 & 0.97   & .333      &        \\
GCC7        & 0.32      & 0.50 & 0.65   & .515      &        \\
GCC8        & 0.25      & 0.50 & 0.51   & .613      &        \\
CPGC        & -0.28     & 1.02 & -0.27  & .784      &        \\
G1GC        & -0.14     & 0.53 & -0.26  & .795      &        \\
PGC         & 0.53      & 0.50 & 1.08   & .282      &        \\
SGC         & -0.17     & 0.50 & -0.34  & .731      &        \\
concurrency & 5.40      & 0.59 & 9.20   & .000      & ***    \\
database    & 4.02      & 0.77 & 5.24   & .000      & ***    \\
functional  & 8.95      & 0.50 & 17.97  & .000      & ***    \\
scala       & 5.95      & 0.50 & 11.95  & .000      & ***    \\
web         & -6.90     & 0.68 & -10.12 & .000      & ***    \\
  \bottomrule
  \end{tabular}
\end{table}

\begin{table}[ht]
  \centering
  \small
  \caption{Hierarchical linear regression model of performance difference ratios between interpretive and tiered execution.}
  \label{tab:Reg-INTvHS-Categories}
  \begin{tabular}{lrrrrrr}
  \toprule   
\textbf{Coefficients} & \textbf{B}        & \textbf{S.E}       & \textbf{t}     &       & \textbf{R2}    & \textbf{p}         \\
\midrule
(Intercept)  & 11.56    & 1.683     & 6.867 & ***   & 0.101 & 0.184     \\
sync         & 3.64E-08 & 2.63E-08  & 1.385 &       &       &           \\
\midrule
(Intercept)  & 11.53    & 1.671     & 6.900 & ***   & 0.112 & 0.162     \\
wait         & 3.77E-04 & 2.58E-04  & 1.461 &       &       &           \\
\midrule
(Intercept)  & 11.81    & 1.794     & 6.581 & ***   & 0.034 & 0.453     \\
notify       & 2.11E-05 & 2.75E-05  & 0.768 &       &       &           \\
\midrule
(Intercept)  & 10.41    & 1.664     & 6.255 & ***   & 0.248 & 0.030     \\
atom         & 7.02E-08 & 2.97E-08  & 2.365 & *     &       &           \\
\midrule
(Intercept)  & 11.82    & 1.643     & 7.191 & ***   & 0.09  & 0.213     \\
park         & 1.67E-06 & 1.29E-06  & 1.294 &       &       &           \\
\midrule
(Intercept)  & 3.14839  & 3.016     & 1.044 &       & 0.402 & 0.004     \\
cpu          & 1.79E-01 & 5.30E-02  & 3.382 & **    &       &           \\
\midrule
(Intercept)  & 12.09    & 1.681     & 7.194 & ***   & 0.031 & 0.468     \\
cache        & 5.18E-18 & 6.977E-18 & 0.742 &       &       &           \\
\midrule
(Intercept)  & 11.17    & 2.049     & 5.45  & ***   & 0.052 & 0.350     \\
object       & 5.27E-09 & 5.48E-09  & 0.962 &       &       &           \\
\midrule
(Intercept)  & 11.60    & 1.906     & 6.085 & ***   & 0.036 & 0.440     \\
array        & 2.94E-08 & 3.72E-08  & 0.791 &       &       &           \\
\midrule
(Intercept)  & 10.20    & 2.079     & 4.908 & ***   & 0.127 & 0.135     \\
method       & 4.60E-10 & 2.92E-10  & 1.571 &       &       &           \\
\midrule
(Intercept)  & 12.35    & 1.806     & 6.839 & ***   & 0     & 0.972     \\
idynamic     & 7.37E-09 & 2.08E-07  & 0.035 &       &       &           \\
  \bottomrule
  \end{tabular}
\end{table}

\section{Analysis}\label{sec5}
In this report, we have presented the results from an analysis of the performance of OpenJDK JRE JVMs executing bytecode in interpreter mode and tiered execution. The study was relative to a contemporary set of benchmark applications. Overall, the results suggest a widening of the performance gap between interpretive execution and tiered execution compared to previous results reported within the literature. For example, early reports of the gap between both modes of execution suggested that the dynamic use of just-in-time compilers increases performance, relative to interpretive execution, by approximately up to 9 times, as reported by Wentworth et al. \cite{Wentworth}. According to the latest findings, there has been a significant increase in the efficiency gap between tiered and interpretive execution, with recent results indicating that the gap is now approximately 20 times greater in favour of tiered execution. Our most recent performance gap analysis confirms this trend, revealing that the efficiency gap between tiered and interpretive execution has widened to 34 times.

Similar to the results presented in previous work that focused on specific workload types, our efficiency observations suggest that workload type significantly affects the performance gap between tiered execution and interpretive execution. The effects from workload type should not be a surprise as tiered execution efficiency, mainly the just-in-time HotSpot compilation undertaken by the C1 and C2 compilers and their subsequent impact is determined by the degree to which the underlying code exhibits specific optimisable patterns. 

When taking a broad view, it is evident from the Renaissance Benchmark workloads that the interpretive and tiered execution approaches have the narrowest difference in performance for the two web-based applications, while the functional and scala categories have the greatest disparity. Furthermore, we have demonstrated statistically significant variations in median performance among the different workload categories.

Are there other explanatory variables? In particular, does the OpenJDK JRE version impact the difference in performance between interpretive and tiered execution? Our results have shown that ignoring workload categorisation: there are no differences in median performance ratios between OpenJDK JRE versions. Our results have shown that focusing specifically on workload category type: the only significant differences, based on the OpenJDK JRE version, are for the database category of application workloads. Median performance ratios increase across OpenJDK JRE versions, from oldest to most recent. The median performance increase indicates that the gap between tiered execution and interpretive execution has widened across OpenJDK JRE versions.

Considering other differentiating characteristics, our results show that CPU usage and the number of atomic operations are the best predictors regarding the difference in the performance between interpreted and tiered execution. 

Our analysis also focused on the effect of changing the build toolchain compiler on OpenJDK JRE performance. Overall, we found no evidence to suggest that changing the GNU GCC compiler version in the OpenJDK JRE build toolchain has any impact. Differences in performance ratios were evident between median workload categories within each GNU GCC version level (all $p < .001$).

Focusing solely on alternating garbage collector's effect on individual OpenJDK JRE performance, our results found no statistically significant differences in performance ratios. This finding raises several questions concerning the out-of-box tuning efficiency of the OpenJDK JRE JVM, particularly the efficiency of the out-of-box JVM for dealing with a wide range of differing workloads. As expected, the serial collector, designed for single-core architectures, has the smallest median performance difference ratio while managing heap memory on a multi-core architecture, irrespective of the OpenJDK JRE version. 

The advantage of running on a multi-core system, a headline characteristic associated with the garbage first G1 collector, was not evident. The literature on garbage collection algorithms generally focuses on metrics such as throughput, latency, and memory footprint, comparing algorithms to algorithms across those metrics domains. Our analysis kept the maximum memory footprint constant across all OpenJDK JRE JVM instances, setting the maximum heap at 2GB. In addition, our research is relative to execution time differences between interpreted and tiered execution. In this work, we have presented the results from a collection of current workloads; these results show that garbage collector choice, possibly except for the serial collector, has no impact on performance differences between interpretive and tiered execution. 

Our results also show that regressing the OpenJDK JRE version, GNU GCC compiler version, garbage collector used to manage heap memory, and workload category on performance difference ratios results in a statistically significant model. In particular, the overall model accounts for 23\% of the variation in performance ratios. The only statistically significant predictor was the workload category. This result would support the strategy of warming up JVMs with specific workloads exhibiting common characteristics and parking those warm JVM instances in wait for similar workloads.

Finally, looking at an alternative workload categorisation strategy, for example, that presented by Prokopec et al. \cite{Renaissance}, that captures the magnitude of operations associated with concurrency and parallel workloads, identified two important characterising metrics. Those metrics were CPU usage and the number of atomic operations. Both metrics account for 40\% and 25\% of the variation in performance difference ratios. Similar to the previous finding, but at a finer granularity, workload characteristics have significant predictive power.  

\section{Conclusion}\label{sec6}
The benefits of just-in-time compilation over traditional interpretive execution in regards to application execution time are widely acknowledged. However, the contemporary literature is absent regarding an investigation into the performance gap between these two models of execution relative to modern concurrent, parallel workloads executing on modern multi-core architectures. In this paper we reposition the performance gap relative to such workloads. Specifically, we analysed the performance gap between JRE JVM interpretive execution relative to JRE JVM tiered execution, considering factors such as JRE version, build toolchain compiler versions, and garbage collection algorithm used to manage heap memory. In addition, we consider workload classification and its predictive power in understanding the performance gap.

Our results show that OpenJDK JRE version, the GNU GCC compiler version used within its build toolchain, or the garbage collection algorithm used for heap memory management have little to no effect in regards to the performance difference between JVM interpretive and JVM tiered execution. However, workload characteristics do have a statistically significant impact on the performance gap between interpretive and tiered execution.

\section*{Credit}
  Conceptualisation, J.L.; methodology,  J.L., R.M. and K.C.; validation,  J.L.;  formal analysis,  J.L.; investigation, J.L.; resources, J.L., R.M. and K.C.; data curation, J.L.; writing---original draft preparation, J.L.; writing---review and editing, J.L., R.M. and K.C.; visualisation, J.L. and K.C.; supervision, R.M. and K.C.; project administration, R.M. and K.C. All authors have read and agreed to the published version of the~manuscript.

\section*{Acknowledgement}
  Jonathan Lambert's PhD research at Maynooth University is funded by the National College of Ireland (NCI) under the NCI Educational Assistance Programme.

\section*{Opendata}
  Sample data are available at the repository: \\ \\
  \url{https://github.com/jonathan-lambert/Interpreter-Tiered-Execution-Data-Paper-arXiv}. \\ 
  (Accessed: April 2023)

\section*{Conflicts}
    The authors declare no conflict of~interest.
    
\bibliographystyle{ieeetr}
\bibliography{references}

\end{document}